# Inferring school district learning modalities during the COVID-19 pandemic with a hidden Markov model

Mark J. Panaggio[1], Mike Fang[1], Hyunseung Bang[1], Paige A. Armstrong[2], Alison M. Binder[2], Julian E. Grass[2], Jake Magid[3], Marc Papazian[3], Carrie K Shapiro-Mendoza[2], Sharyn E. Parks[2]

[1] Johns Hopkins University Applied Physics Laboratory, Laurel, Maryland

[2] Palantir Technologies, Denver, Colorado

[3] CDC COVID-19 Response Team

## Abstract

In this study, learning modalities offered by public schools across the United States were investigated to track changes in the proportion of schools offering fully in-person, hybrid and fully remote learning over time. Learning modalities from 14,688 unique school districts from September 2020 to June 2021 were reported by Burbio, MCH Strategic Data, the American Enterprise Institute's Return to Learn Tracker and individual state dashboards. A model was needed to combine and deconflict these data to provide a more complete description of modalities nationwide.

A hidden Markov model (HMM) was used to infer the most likely learning modality for each district on a weekly basis. This method yielded higher spatiotemporal coverage than any individual data source and higher agreement with three of the four data sources than any other single source. The model output revealed that the percentage of districts offering fully in-person learning rose from 40.3% in September 2020 to 54.7% in June of 2021 with increases across 45 states and in both urban and rural districts. This type of probabilistic model can serve as a tool for fusion of incomplete and contradictory data sources in support of public health surveillance and research efforts.

## Introduction

During the COVID-19 pandemic, non-pharmaceutical interventions are among the primary tools for mitigating the transmission of the virus. Even with global vaccination efforts, the emergence of variants and waning effectiveness of vaccines mean that policies for encouraging and even requiring physical distancing, handwashing, and mask wearing continue to be implemented around the world. Congregate settings, such as schools, residential facilities, and correctional institutional settings present challenges for implementing certain measures, such as physical distancing. In other respiratory disease outbreaks, such as influenza, schools are believed to play a key role in transmission, and school closures are often recommended as a means for controlling outbreaks [1], [2]. In March 2020, schools around the United States were closed for full-time in-person learning and many remained closed until the end of the school year, implementing remote learning instead.

By late summer 2020, with the pandemic showing no signs of subsiding, schools across the U.S. from kindergarten through 12th grade (K–12) faced the difficult decision of whether to reopen for in-person learning or to adopt alternative strategies [3]. There were widespread concerns that reopening schools prematurely could exacerbate COVID-19 transmission and create substantial public health risks, while reductions in face-to-face instruction caused by school closures would negatively impact learning as well as the mental and physical well-being of students [4]-[6]. Consequently, schools adopted a range of strategies from fully remote to fully in-person to balance concerns about safety and educational quality.

Over the course of the 2020–2021 school year, evidence accumulated that the reduction in face-to-face instruction and the adoption of new learning approaches was leading to educational gaps, particularly for those without reliable access to the Internet [4]-[6]. The learning losses caused by this disruption to the education system could be life-long in these students, leading to lower future earning potential, productivity, and even decreases in gross domestic product for the entire country [7].

By January 2021, several studies found that there was "little evidence that schools have contributed meaningfully to increased community transmission" when other mitigation strategies are in place [3], [8]. In response, there was an executive order calling for a return to "safe, in-person learning as quickly as possible" [9] and resources were mobilized to provide aid to students and educators [10]. However, a coordinated and strategic nationwide approach to re-opening schools requires accurate information on learning modality by school district and timely updates of offerings over time. Although no national database for tracking these modalities was available, some school districts and several states reported this information through public-facing online dashboards. Small-scale information collection was being done by a few private companies prior to the pandemic: American Enterprise Institute/Return 2 Learn Tracker (data collected from website and online dashboard scraping) [11], Burbio (data collected from media reports and website scraping) [12], and MCH Strategic Data (data collected from phone surveys) [13]. On their own, these diverse data sources provided an incomplete picture of learning modality offerings due to limited coverage and missing dates, and some data sources were discrepant. A tool for fusing these data sources and resolving these conflicts was needed.

In response, we implemented a hidden Markov model (HMM) [14], [15] that leverages historical data on the learning modalities offered during the 2020–2021 school year from each of these sources to infer the most likely modalities offered by over 14,000 public school districts nationwide. Here, we discuss the implementation of this model and how it provided real-time data for critical decision-making and reopening schools nationwide.

## Methods

*Data sources and inclusion criteria*

We collected learning modality data from four different sources and aggregated data by week (Table 1). These sources reported learning modalities by school district collected using a variety of methods including automated and manual data collection from websites and dashboards, systematic monitoring of media reports, and phone surveys of school and district administrator offices. All traditional and charter public K–12 school districts open during the 2020–2021 school year were eligible for inclusion except for those whose names included the words 'online', 'cyber', 'distance', 'remote', 'virtual', or 'digital' as an indication of operating status prior to the pandemic. These were excluded because they generally do not offer traditional in-person learning even under non-pandemic circumstances. Open school districts were defined as local education agencies with types 1 (regular public school district that is a component of a supervisory union), 2 (regular public school district that is not a component of a supervisory union) or 7 (independent charter districts) with school year statuses 1 (Open), 3 (New), 4 (Added), 5 (Changed), and 8 (Reopened) as categorized by the National Center for Education Statistics [16]. Together, the data sources represented learning modalities from 14,688 districts that encompass 84.5% of the 17,375 eligible districts across the country during the 2020–2021 school year. These districts enrolled approximately 48 million students and included over 57% of students from eligible districts in each of the 50 states [16].

Table 1: *Learning modalities reported by* source. The states *column includes the District of Columbia. State dashboards were available from CO [17], CT [18], HI [19], ID [20], IL [21], LA [22], MN [23], MO [24], NC [25], NM [26], OH [27], OR [28], SC [29], TN [30], VA [31], VT [32], and WA [33].*

| Source | Districts (no.) | District-weeks (no.) | States (no.) | Data collection method | Update Pattern |
|---|---|---|---|---|---|
| Burbio | 1,187 | 37,589 | 51 | Media reports, website scraping | Weekly |
| MCH Strategic data (MCH) | 13,975 | 58,137 | 51 | Phone surveys | Irregular |
| Return to learn tracker (R2LT) | 8,601 | 343,596 | 51 | Website and dashboard scraping | Weekly |
| State Dashboards (SD) | 4,007 | 24,732 | 17 | Dashboard scraping | Daily |
| Total (unique) | 14,688 | 379,726 | | | |
| HMM (overall) | 14,688 | 616,896 | 51 | Inferred modalities using HMM | Weekly |
| HMM (High confidence) | 13,275 | 463,946 | 51 | Inferred modalities using HMM (probability ≥0.75) | Weekly |

*Learning modality definitions*

The learning modalities offered by school districts nationwide were divided into three categories: **full-time in-person**: districts that offered face-to-face instruction five days per week to all students at all grade levels; **full-time remote**: districts that did not offer any face-to-face instruction; **hybrid**: all other districts. The hybrid category encompassed districts that offer face-to-face instruction to some, but not all grades or schools in the district, as well as those that offer face-to-face instruction to all grades or schools less than five days per week. These definitions were agreed upon as the common categories used by the federal agencies most closely tracking learning modalities during the pandemic [11]–[13].

*Description of hidden Markov model*

An HMM is a sequential model that assumes that the system evolves according to a random process. HMMs have been successfully used for prediction and inference problems across a variety of domains including data analysis, speech recognition, and bioinformatics [34]. In this formulation, the probability distribution for the next state is completely determined by the current state and can be completely described by an $n \times n$ *transition matrix* where $n$ is the number of possible states. In this context, the hidden *state variable* represents the true learning modality and the transition matrix describes the probability of transitioning between each pair of learning modalities. In an HMM, one cannot observe the state variable directly, but instead one observes a collection of *emission variables* whose probability distributions are dependent on this hidden state. We interpreted the learning modalities reported by the four sources as emission variables that are noisy versions of the hidden state. In other words, we assumed that there were unknown nonzero probabilities that each source would report incorrect modalities. When determining the true modality most likely to generate the observed data, sources with a higher probability of reporting the correct modality would have a greater influence when inferring the true modality from the observed data.

The choice of an HMM was motivated by the observation that the learning modalities offered by each district tended to be consistent over time, so knowing the current modality provides information about the most likely modality for the previous week and the next week. This implies that one should consider the entire sequence of observed learning modalities for a given district when attempting to infer the most likely values for any missing modalities.

We numbered the learning modalities from one to three corresponding to fully remote (1), hybrid (2) and fully in-person learning (3), respectively. In this formulation, there are a total of 45 parameters, including a $3 \times 3$ matrix of transition probabilities $T$ where $T_{ij}$ denotes the probability of transitioning from modality $i$ to modality $j$, and four $3 \times 3$ matrices of emission probabilities $B$, $M$, $R$, and $S$ corresponding to the four sources: Burbio, MCH, Return to Learn Tracker (R2LT), and State Dashboards (SD) respectively. Here $B_{ij}$ represents the probability that Burbio will report modality $j$ for a district with true modality $i$. The components of $M$, $R$ and $S$ are defined in a similar manner.

Given these parameter values, one can use maximum likelihood estimation to ascertain the sequence of true learning modalities that is most likely to produce the reported modalities for each district. In practice those parameters are unknown, but a method known as the Baum-Welch algorithm can be used to estimate them and the hidden states simultaneously [14]. This method is a special case of the expectation maximization algorithm [35], in which one alternates between updating estimates of the parameters and hidden states in such a way that the likelihood of the observed data increases monotonically until reaching a local maximum. This estimation process was carried out using a python package called pomegranate [36].

Because the true modalities are unknown and the reported modalities from each source may be incorrect, this model was trained in an unsupervised manner. The hidden states inferred by the model correspond to clusters of observations that were then mapped to learning modalities based on the most common modality reported within each cluster. These inferred states are based on the consensus of available data, so any biases present in these data will be reflected in biases in the inferred modalities.

*Modality agreement*

The goal of the HMM is to provide the most reliable prediction of learning modality by week for each district using multiple data sources. We defined agreement as consistency between the reported modalities of different sources in the same district and week. The proportion of district-weeks that each individual source agreed with the HMM was compared to the proportion of district-weeks that each individual source agreed with the remaining sources. A one-sided Student's t-test was used to determine whether the HMM provided higher agreement than the other sources [37].

# Results

*HMM Performance*

The parameters of the HMM were estimated using the most current data available for that week during September 1, 2020, and June 25, 2021. The estimated parameters are displayed in Table 2 and Table 3. The model predicted changes in learning modality relatively rarely, with fully in-person learning remaining in-person 98% of the time, hybrid learning remaining hybrid 96% of the time, and fully remote learning remaining remote 90% of the time. Also, the model predicted that reported modalities agree with the true modality most of the time. According to the model, R2LT was most consistent with the inferred modality, inferring modality (99%, 99%, 99%) of the time for remote, hybrid and in-person learning, respectively. For SD, Burbio, and MCH, those quantities were (64%, 86%, 96%), (81%, 62%, 82%) and (80%, 78%, 60%), respectively.

*Table 2: Transition matrix for HMM. Rows correspond to the current modality, columns correspond to the next modality and values represent probabilities of transitioning between these states. This matrix was estimated directly from data.*

|  |  | NEXT MODALITY | | |
| --- | --- | --- | --- | --- |
|  |  | Remote | Hybrid | In-person |
| **CURRENT MODALITY** | Remote | 0.903 | 0.079 | 0.018 |
|  | Hybrid | 0.014 | 0.961 | 0.025 |
|  | In-person | 0.004 | 0.013 | 0.983 |

*Table 3: Emission probabilities. Rows correspond to the true modality, columns correspond to the reported modality, and values represent the estimated probabilities that each source would report that modality correctly.*

| SOURCE | | BURBIO | | | MCH | | | R2LT | | | SD | | |
|---|---|---|---|---|---|---|---|---|---|---|---|---|---|
| REPORTED MODALITY | | Remote | Hybrid | In-person | Remote | Hybrid | In-person | Remote | Hybrid | In-person | Remote | Hybrid | In-person |
| TRUE MODALITY | Remote | 0.805 | 0.145 | 0.050 | 0.795 | 0.178 | 0.027 | 0.992 | 0.008 | 0.001 | 0.642 | 0.330 | 0.028 |
| | Hybrid | 0.067 | 0.617 | 0.317 | 0.090 | 0.775 | 0.135 | 0.002 | 0.997 | 0.001 | 0.003 | 0.863 | 0.134 |
| | In-person | 0.016 | 0.161 | 0.823 | 0.055 | 0.347 | 0.598 | 0.001 | 0.001 | 0.997 | 0.001 | 0.042 | 0.956 |

We found that the inferred modalities show high agreement with the reported modalities from all data sources across the period (). The modalities inferred by the HMM had higher agreement with R2LT (99.6%) and SD (90.1%) than with MCH (69.3%) and Burbio (70.7%). In pairwise comparisons, we found that SD, R2LT and MCH had a higher agreement with the HMM than with any other source ($P = 0.038$ for MCH and SD and $P <$ 0.01 for all other comparisons, e). Agreement between Burbio and the HMM was also high, but was surpassed by agreement between Burbio and SD (80.0%).

The proportion of weeks in agreement and number of districts covered by each pair of sources was generally consistent over the period of the study (). However, the number of districts where multiple sources were available varied substantially due to many state dashboards not being available prior to February 2021. The MCH survey was also inactive during February and March of 2021 and had a variable response rate.

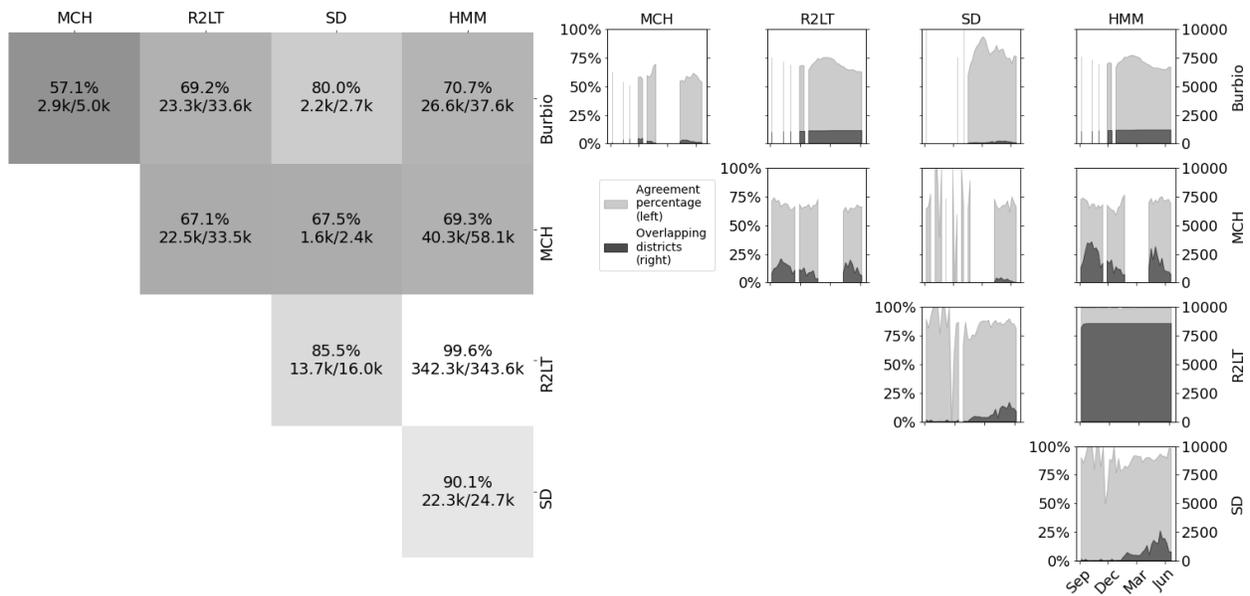

*Figure 1: Pairwise agreement in reported modalities by source. The left panel shows the percentage and total number of district-weeks where modalities were reported by each pair of sources and where those modalities matched. The right panel displays the agreement percentage (light, left y-axis) and number of districts where both sources were available (dark, right y-axis) over time.*

*National trends in learning modalities*

Using the HMM, we were able to provide a comprehensive view of the learning modalities offered nationwide. We found a steady increase in districts reporting full time in-person learning from 40.3% in first week of September 2020 to 54.7% at the last week of June 2021 (Figure 2). The proportion of full-time in-person learning steadily increased throughout the period except during November–December, 2020, which coincided with a spike in COVID-19 cases nationally. The percentage of districts offering only remote learning was less than 5% by the week of March 14, 2021 and had decreased to 2% by the last week of June 2021, which coincided with the peak of the rollout of vaccines [38]. Overall, a move towards full-time in-person learning was observed during the study period.

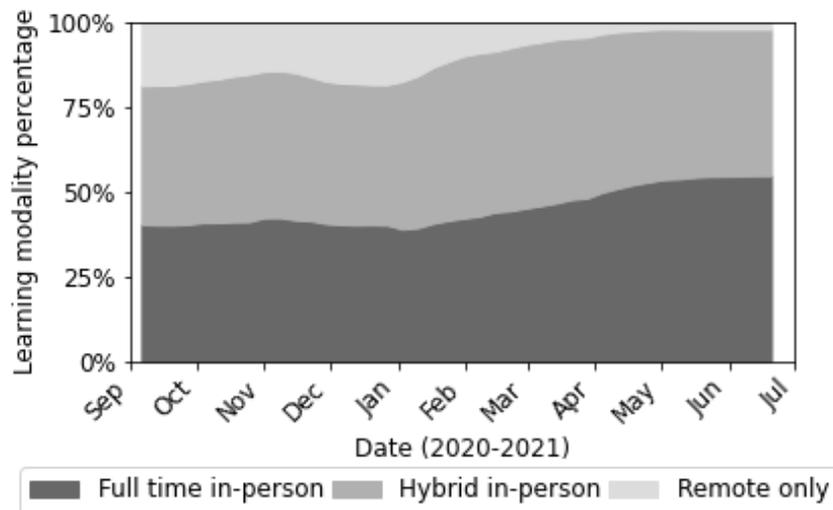

*Figure 2: Percentage of U.S. school districts in each modality. Here the shaded bands represent the share of district offering in-person (dark), remote (light) and hybrid learning modalities (medium).*

*Geographic trends in learning modalities*

The percentage of districts offering in-person learning during three time periods, the first week of September 2020, the first week of January 2021, and the last week of June 2021 are shown in *Figure 3*. Florida was the state with the highest proportion of full time in-person learning districts, fully reopening by the end of June 2021 (100.0%). Other states with a high proportion of full-time in-person learning include South Carolina (97.5%) and Iowa (95.7%). States and districts with the lowest proportion of full-time in-person learning at the end of the study period included Louisiana (8.4%), Vermont (10.5%), and Delaware (15.0%). States with the greatest increase in in-person learning during September 2020–June 2021 included South Carolina (from 6.3% to 97.5%) and Alabama (from 10.7% to 77.1%). Note that Hawaiian schools operate under a single district that offered hybrid learning throughout the entire study period according to available data and the model.

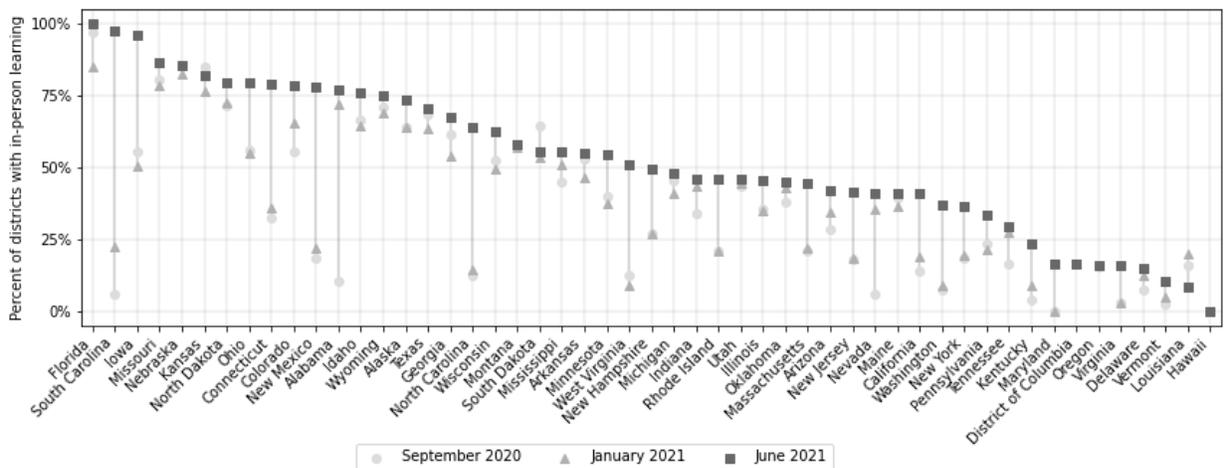

*Figure 3: Estimated in-person learning percentages. Points indicate the percentages of districts offering full time in-person learning on the following dates: September 20, 2020 (circle); January 21, 2021 (triangle); and June 21, 2021 (square).*

*Learning modalities by urban-rural classification*

To provide insight into which areas and populations were most affected by access to in-person learning, each school district was mapped to a county using the address reported by the National Center for Education Statistics and then categorized according to the National

Center for Health Statistics' six urban-rural classifications: large central metro, large fringe metro, medium metro, small metro, micropolitan, and non-core areas [39]. Although the overall trends mirror those for the entire United States, we found that remote learning was most prevalent in areas categorized as large central metropolitan areas and that rural districts and small metropolitan areas had the highest proportion of districts offering full-time in-person learning (Figure 4). Furthermore, the increases in the percentage of full-time in-person districts over the study period (first week of September 2020 vs. last week of June 2021) were the greatest among the large fringe metro (28.6% to 48.4%), followed by medium metro (30.7% to 49.8%) and small metro areas (43.6% to 60.4%), and smallest among large core metro (20.7% to 32.5%), micropolitan (47.4% to 60.5%), and non-core (59.3% to 68.1%).

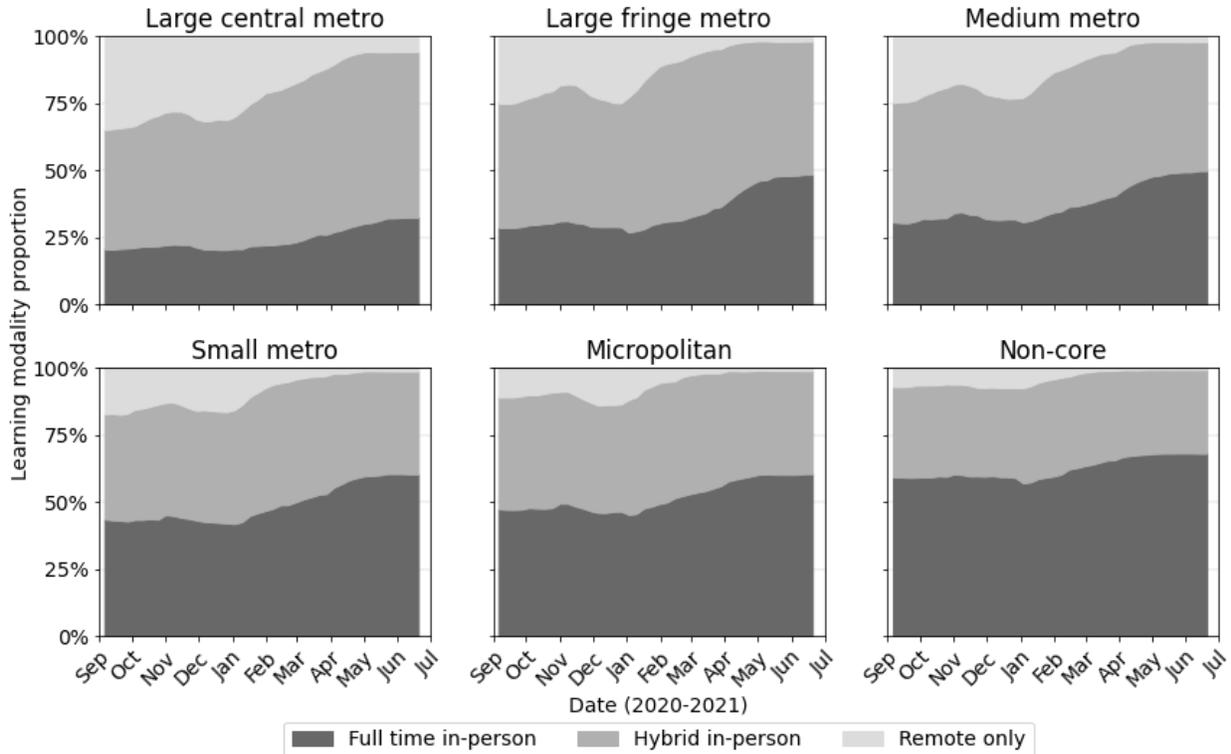

*Figure 4: Percentage of U.S. school districts in each modality by urban/rural classification. Urban-rural classifications are based on the county where each school district is located and are defined by the National Center for Health Statistics [39]. Here the shaded bands represent the share of district offering in-person (dark), remote (light) and hybrid learning modalities (medium).*

## Discussion

The hidden Markov Model and inferred modalities presented here provide several clear advantages over the raw modality data from Burbio, MCH, R2LT and SD individually. By leveraging multiple sources and inferring modalities in weeks where data are not available, our approach provides greater geographic and temporal coverage than any single data source. It provides a principled method for resolving conflicts between the reported modalities without relying on arbitrary source prioritization schemes by learning the weights (emission probabilities) to assign to the reported modalities from each source that maximize the likelihood of the available data. Subject matter experts from multiple federal agencies assessed each data source and weights were assigned commensurate with the determined hierarchy. This approach also makes it possible to ascertain consensus modalities in weeks where no data are available and provides confidence estimates for each modality to allow end-users to determine where the stated modalities are uncertain. These estimates agree with data sources where data are available.

Our analysis has several limitations. First, it is limited by the quality of the available data. No ground truth data were available to fully validate the accuracy of the inferred modalities, which represent a consensus of several sources, which may be incorrect if systematic biases are present. Inconsistencies in the way these sources defined learning modalities is one potential source of bias. For example, state dashboards occasionally reported modalities by school or grade and did not always report modalities using the same three categories. In these cases, the reported policies were mapped to district level policies in agreement with the definitions above. Second, the districts included are those for which modalities were reported by the Burbio, MCH, R2LT and SD. Districts without reported modalities tended to be smaller. Overall, the median enrollment and number of schools per district were 396 and 1, respectively, as compared to 1080 and 3 for eligible districts. The districts in each learning modality may therefore not be representative of all districts nationwide. Third, the HMM is based on the Markov assumption that the probability distribution for modalities next week depends only on the modality currently offered. This assumption may be violated in instances where schools closed temporarily due to an outbreak with the intent of reopening once the outbreak was over.

## Conclusions

The HMM can serve as a valuable tool for monitoring school reopening and providing situational awareness during public health responses. Using our model, we were able to infer the modalities offered in over 14,000 districts throughout the 2020–2021 school year. We observed decreases in full-time in-person learning between September 2020 and January 2021 in 20 states, which coincided with the peak incidence of COVID-19 cases in the United States that year. This was followed by steady increases in the number of school districts offering full-time in-person learning starting in January 2021. We also found that school districts in rural areas were more likely to offer full-time in-person learning than those in large metropolitan areas. Both observations are consistent with media reports and anecdotal evidence. These data could serve as a valuable tool for evaluating the impact of school policies on COVID-19 transmission, educational attainment and other outcomes.

Starting in April 2021, these findings were shared with public health officials within the U.S. Centers for Disease Control, the U.S. Department of Education and the White House via a weekly Federal Interagency School Data Report. During the 2021–2022 school year, this model continues to be used to provide situational awareness as schools respond to the ever-changing conditions of the COVID-19 pandemic [40].

## Acknowledgments

This work was supported by the US Department of Health and Human Services Office of the Assistant Secretary for Preparedness and Response Contract [contract number 75A50121C00003]. The authors would like to thank Burbio, MCH strategic data and the American Enterprise Institute for providing data used in this study as well as Nick Simmons (Department of Education) for his support and advice during this project.

## Disclaimer

The findings and conclusions in this report are those of the authors and do not necessarily represent the views of the US Centers for Disease Control and Prevention. This activity was reviewed by CDC and was conducted consistent with applicable federal law and CDC policy (see 45 C.F.R. part 46; 21 C.F.R. part 56).

[3] Honein MA, Barrios LC, Brooks JT. Data and Policy to Guide Opening Schools Safely to Limit the Spread of SARS-CoV-2 Infection. *JAMA* 2021; 325:**9**:823–824, doi: [10.1001/jama.2021.0374](10.1001/jama.2021.0374).

[4] Schleicher A. The impact of covid-19 on education insights from education at a glance 2020. *oecd.org* 2020; [https://www.oecd.org/education/the-impact-of-covid-19-on-education-insights-education-at-a-glance-2020.pdf](https://www.oecd.org/education/the-impact-of-covid-19-on-education-insights-education-at-a-glance-2020.pdf)

[5] Pokhrel S, Chhetri R. A literature review on impact of covid-19 pandemic on teaching and learning. *High Educ Fut 2021;* 8:**1**:133–141.

[6] Dhawan S. Online learning: A panacea in the time of covid-19 crisis. *J Educ Technol Syst* 2020;. 49:**1**:5–22.

[7] Hanushek EA, Woessmann L. The economic impacts of learning losses. OECD Educ Working Papers 2020; 225. doi: [https://doi.org/https://doi.org/10.1787/21908d74-en](https://doi.org/https://doi.org/10.1787/21908d74-en).

[8] COVID-19 in children and the role of school settings in covid-19 transmission. *Stockholm: ECDC* 2020. https://www.ecdc.europa.eu/en/publications-data/children-and-school-settings-covid-19-transmission

[9] Supporting the Reopening and Continuing Operation of Schools and Early Childhood Education Providers. Executive order. 140000: 7215–7218. [https://www.federalregister.gov/documents/2021/01/26/2021-01864/supporting-the-reopening-and-continuing-operation-of-schools-and-early-childhood-education-providers](https://www.federalregister.gov/documents/2021/01/26/2021-01864/supporting-the-reopening-and-continuing-operation-of-schools-and-early-childhood-education-providers)

[10] 100 days of the Biden administration: How the department of education has helped more schools safely reopen and meet students' needs. *[www.ed.gov](www.ed.gov)* 2021;
20